# Nucleation of crystal surfaces with corner energy regularization

T. Philippe*, H. Henry, M. Plapp

Physique de la Matière Condensée, Ecole Polytechnique, CNRS, Université Paris-Saclay, 91128 Palaiseau, France

Abstract: The thermodynamics of strongly anisotropic crystalline surfaces is analogous to that of a binary mixture exhibiting phase separation. On a metastable planar surface, formation of stable orientations requires a nucleation process, in which the energy associated with the presence of corners must be considered. In this context, a nucleation event corresponds to the formation of a critical shape for the crystalline surface before the system enters the growth regime. We first derive the Euler-Lagrange equation for crystal surface nucleation, in two dimensions, and show that the saddle-point condition corresponds to a vanishing chemical potential along this critical surface. We then perform numerical simulation of the equation of motion for the crystal surface and show that, as compared with saddle point nucleation, ridge crossing is dynamically favoured.

* Corresponding author: thomas.philippe@polytechnique.edu

## I. INTRODUCTION

The problem of determining the equilibrium shape of a crystallite is well understood in the sharp-interface framework, in which the crystal surface is assimilated to a smooth line and atomistic details such as steps and adatoms are not considered. This shape is spherical (or circular in two dimensions) in the absence of anisotropy. If the surface free energy is anisotropic, the shape deviates from a sphere: low-energy directions are preferred and occupy larger surface area. If the anisotropy is sufficiently strong, edges and corners are formed, and the directions with the highest energies are excluded from the equilibrium shape. This problem has been studied for over a century by many authors [1-9]. The Wulff construction links the polar plot of the surface free energy $\gamma(\theta)$ (where $\theta$ is the tangent angle) to the equilibrium shape of the crystal and dates to 1901 [3]. Subsequently, many elegant descriptions of the equilibrium shape of a crystalline surface were proposed by Herring [1,2], Burton [4] and Frank [5], by Cabrera with the double tangent construction [6,7], by Cahn and Hoffman who introduced the ξ-vector [8], and by Andreev [9].

It is useful to recall a few concepts that can be illustrated with the help of the polar plot of the Cahn-Hoffman ξ-vector, see Fig.1a. For weak anisotropies, this plot is convex and gives directly the equilibrium shape. As soon as, for some orientations, the interface stiffness ($\gamma + \gamma_{\theta\theta}$ with $\gamma_{\theta\theta}$ the second derivative of $\gamma$ with respect to $\theta$) becomes negative, the plot develops self-intersections and turning points. The equilibrium shape is now given by the inner envelope and exhibits corners at the self-intersection points. The orientations located in between the two "flanks" of the corner are missing from the equilibrium shape. These directions form the so-called "ears" on the ξ-plot. Surfaces with orientations that are located between the self-intersection and the turning point are metastable, that is, they are stable against infinitesimal fluctuations of the orientation, but decompose into the two stable directions at the corner point if a perturbation of finite amplitude is provided. In contrast, directions located between the two turning points have a negative stiffness and are unstable to wrinkling, with a growth rate that diverges when the typical wavelength of the wrinkles tends to zero. This implies that dynamic equations for the growth of such surfaces are ill-posed since they are backward parabolic.

To circumvent this issue, the problem can be regularized by adding a curvature-dependent term to the surface energy, as in [10]. This additional term introduces a new length scale, over which the sharp corners are rounded. The curvature-dependent term also smoothes out the small-scale instability. It was shown by Spencer [11] that the regularized sharp interface model converges toward the Wulff shape when the regularization parameter approaches zero. The results of the asymptotic analysis performed by Spencer [11] reveal that such a regularization procedure is physically sound. Its effect on the equilibrium shape is well established, at least in two dimensions; the generalization to three dimensions is non-trivial.

A close analogy can be established with the thermodynamics of phase separation in a binary mixture. Here, the orientation $\theta$ is analogous to the composition c of the mixture, and the surface stiffness to the second derivative of the free energy. The two coexisting phases are equivalent to the two orientations on the two sides of a corner. In the mixture, a supersaturated phase is metastable as long as the free energy as a function of composition is convex; the equilibrium phase appears through a nucleation process. In contrast, for a concave free energy, spinodal decomposition occurs, that is, spontaneous amplification of microscopic composition fluctuations. In this analogy, the curvature regularization corresponds to the introduction of the square concentration gradient term in the Cahn-Hilliard theory.

Equilibrium and dynamics of anisotropic surfaces have also been studied extensively in diffuse-interface (phase-field) theories, in which the interfaces are diffuse and have a well-defined thickness [12]. They have been found to be ill-posed for surfaces with negative stiffness and can be regularized by adding a curvature dependent term in the energy.

Here, we study the case of metastable surfaces, in which a nucleation event is necessary to reach the true equilibrium. Indeed, the effect of regularization on the equilibrium shape and on the dynamics of stable orientation formation in the spinodal regime as well as coarsening either by surface diffusion and motion by curvature or during reaction-limited crystal growth from a melt have been thoroughly studied, both in phase-field models and sharp-interface theories [13-23]. However, the effect of the corner energy on the nucleation regime remains unclear and even though the analogy with Cahn-Hilliard theory has already been shown in the context of spinodal decomposition of facets [19], it is difficult to anticipate, a priori, the shape of the nucleating surface and its dynamic evolution.

The formation of the critical shape that corresponds to the saddle-point of the surface energy, in which the corner energy must be considered, obeys the thermodynamics of nucleation. Once the critical surface is formed, the system enters the growth regime, either by surface diffusion or by an evaporation-condensation mechanism. In this context, the term growth refers to an increase in size of facets and other surface features, at constant volume of the solid phase. The aim of the present paper is to understand the effect of regularization on the critical shape and its properties. We first briefly present the analogy of the corner problem to a phase transition, closely following Spencer [11], and we then formulate the nucleation problem of crystal surface formation and determine the properties of the critical surface, discussing the close analogy with the diffuse-interface theory of nucleation [24]. We then explore dynamics for growth controlled by an evaporation-condensation mechanism (motion by curvature).

## II. A PHASE TRANSITION AT THE INTERFACE

In two dimensions, the solid surface is described by a curve in the ($x,y$) plane which is parametrized by an arclength s chosen so that when travelling in the increasing s direction along the interface, the solid is on the right and the orientation of the interface is given by the tangent angle to the interface $\theta$ measured clockwise from the $x$-axis (see Fig.1b). The total energy of the surface is

$$E = \int \gamma^* ds \quad (1)$$

where the surface energy per unit length is the sum of the surface energy density and of a corner regularization term

$$\gamma^* = \gamma(\theta) + \frac{1}{2}\beta\kappa^2 \quad (2)$$

with $\beta$ the regularization parameter and the local curvature of the surface given by

$$\kappa = \frac{d\theta}{ds} \quad (3)$$

which is positive for a solid bump.

Minimizing the total energy of the surface (Eq.1) subject to a fixed solid area constraint gives the following form of the chemical potential for an anisotropic and regularized surface, expressed as the difference between its actual value and that of a flat interface,

$$\mu = \Gamma(\theta)\kappa - \beta\left(\frac{d^2\kappa}{ds^2} + \frac{1}{2}\kappa^3\right) \quad (4)$$

with the stiffness

$$\Gamma(\theta) = \gamma(\theta) + \gamma''(\theta). \quad (5)$$

For a solid in equilibrium, the chemical potential is constant, and the surface bounds the solid with a prescribed area. In the absence of regularization, the crystal shape is size-independent and the effect of $\mu$ is to modify the length scale: $\mu$ scales inversely with the crystal dimensions. The equilibrium shape of the crystal without regularization has been shown to depend on the details of the surface energy density. If for some orientations the stiffness is negative, the crystal shape contains a corner that truncates a given range of orientations. It was shown by Cabrera [7] that truncating these orientations at equilibrium corresponds to a minimization of the surface energy with a double-tangent construction spanning the range of orientations which are not on the convex envelope of the projected surface energy. The orientations between the two tangent points correspond to the missing orientations at the corner. The equilibrium shape is given by the interior part of the ξ-plot (see Cahn and Hoffman [8] for more details),

$$\xi = \gamma\mathbf{n} + \gamma'\mathbf{t} \quad (6)$$

with **n** and **t** the normal and tangent unit vectors at the surface. The self-intersection points of the ξ-plot correspond to the orientations on either side of the corner, i.e. orientations between these values are missing in the equilibrium crystal shape and form an "ear" in the ξ-plot, see Fig.1a. It is possible to construct a potential from the surface energy that has a double-well shape [7]. The standard common-tangent construction then yields the two corner orientations, and the equilibrium crystal shape corresponds to the convex envelope of this potential. This double-well function, denoted *f* in the following, is defined as the surface energy projected on the *x*-axis

$$E = \int f(q) dx \qquad (7)$$

where *f(q)* is given by

$$f(q) = \frac{\gamma(\theta)}{\cos\theta} \qquad (8)$$

with the surface slope

$$q = \tan\theta. \qquad (9)$$

Cabrera [7] showed that the energy is minimized by the convex envelope of *f(q)*. The convexity of this double well function is given by the sign of the surface stiffness as

$$\frac{d^2 f}{dq^2} = \Gamma(\theta)\cos^3(\theta) \qquad (10)$$

and $cos^3(\theta)$ is positive on $[-\pi/2, \pi/2]$. If the stiffness is positive for all orientations, i.e. *f(q)* is convex, then the equilibrium crystal shape contains all orientations, whereas if for some orientations the stiffness is negative the equilibrium shape omits the orientations for which *f(q)* is nonconvex, replacing the nonconvex region by the common-tangent construction (see Fig.2), and a corner appears on the equilibrium shape. As shown by Spencer [11], the corner orientations *θ⁻* and *θ⁺* on either side of the corner (i.e. where the ξ-vector intersects) satisfy

$$\left[\gamma'(\theta)\cos(\theta) + \gamma(\theta)\sin(\theta)\right]_{\theta^-} = \left[\gamma'(\theta)\cos(\theta) + \gamma(\theta)\sin(\theta)\right]_{\theta^+} \qquad (11)$$

$$\left[-\gamma'(\theta)\sin(\theta)+\gamma(\theta)\cos(\theta)\right]_{\theta^-}=\left[-\gamma'(\theta)\sin(\theta)+\gamma(\theta)\cos(\theta)\right]_{\theta^+} \quad (12)$$

which precisely corresponds to the common-tangent construction

$$\left.\frac{df}{dq}\right|_{q^-}=\left.\frac{df}{dq}\right|_{q^+} \quad (13)$$

$$\left[f-q\frac{df}{dq}\right]_{q^-}=\left[f-q\frac{df}{dq}\right]_{q^+} \quad (14)$$

with $q^-$ and $q^+$ the end points of the common tangent and thus the slopes on either side of the corner of the crystal shape. It is worth mentioning that the corresponding orientations ($\theta^-$ and $\theta^+$) are not the end points of the common tangent of $\gamma(\theta)$ but of $f(q)$. The orientations between $\theta^-$ and $\theta^+$ are metastable with respect to the formation of the stable corner orientations when the curvature of the double-well function is positive, and thus for positive stiffness, and unstable for negative stiffness. In this sense the corner formation obeys the thermodynamics of a first order phase transition. When for some orientations the surface stiffness is negative, formation of stable orientations is closely analogous to the spinodal decomposition of a binary alloy, and this has been thoroughly investigated [22,23]. For metastable orientations, a nucleation event is required during which the regularization term of the energy associated with the presence of corners must be considered.

Using matching asymptotic expansions, the corner shape in the presence of regularization was determined [11], and it was shown that the regularized shape approaches the sharp-interface result when the regularization strength approaches zero. The width of the corner-rounding region scales with $\beta^{1/2}$ with a radius of curvature of order $\beta^{1/2}$ [11]. Finally, the exact solution for the regularized equilibrium shape of a semi-infinite wedge was described by Spencer [11], from which we determine in the next section the critical shape associated with the saddle-point of the nucleation barrier. Indeed, in both situations the chemical potential vanishes.

## III. A THEORY OF NUCLEATION OF CRYSTAL SURFACES

### 1. Derivation of the Euler-Lagrange equation

In the case of nucleation, which is of practical interest since there is, to our knowledge, no analogy to a quench that would abruptly put the system in an unstable region, the initial state is metastable. Here, for the sake of simplicity, we consider a simple semi-infinite two-dimensional crystal with a surface given by:

$$y_0(x) = -q_0 x \quad (15)$$

with a slope $q_0$ that lies in the metastable region $[q^-, q_s]$ with $q_s$ the slope such that

$$\left. \frac{d^2 f}{dq^2} \right|_{q_s} = 0 \quad (16)$$

which is is the inferior limit of the spinodal region (see Fig.2). The regularized energy density depends on both the surface orientation and the surface curvature as in Eq.2. The nucleation barrier is defined as the difference in energy between the energy of the flat initial surface (with the slope $q_0$), denoted $E_0$, and the energy of the critical shape ($E_c$) associated with the saddle point of the total energy as given in Eq.1. Minimization of the total energy is subject to the volume constraint

$$\int y \, dx = \int y_0 \, dx \quad (17)$$

yielding the following Euler-Lagrange equation

$$-\lambda - \frac{d}{dx} \frac{\partial Q}{\partial y_x} + \frac{d^2}{dx^2} \frac{\partial Q}{\partial y_{xx}} = 0 \quad (18)$$

where $\lambda$ is the Lagrange multiplier and with

$$Q = \gamma^* \sqrt{1 + y_x^2} - \lambda(y - y_0). \quad (19)$$

The Euler-Lagrange equation gives

$$-\lambda - y_{xx}\left(1+y_x^2\right)^{-3/2}\left[\gamma + 2\frac{\partial \gamma}{\partial y_x}y_x\left(1+y_x^2\right)+\frac{\partial^2 \gamma}{\partial y_x^2}\left(1+y_x^2\right)^2\right]$$
$$+\beta\left[\left(1+y_x^2\right)^{-5/2}y_{xxxx} - 10\left(1+y_x^2\right)^{-7/2}y_{xxx}y_{xx}y_x + \left(1+y_x^2\right)^{-9/2}\left(15y_{xx}^3 y_x^2 - \frac{5}{2}y_{xx}^3\right)\right] = 0 \quad (20)$$

Using $\kappa = -y_{xx}\left(1+y_x^2\right)^{-3/2}$ and $\gamma_{y_x} = -\gamma_\theta\left(1+y_x^2\right)^{-1}$, this can be rewritten as

$$\lambda = \left(\gamma + \frac{d^2\gamma}{d\theta^2}\right)\kappa - \beta\left(\frac{d^2\kappa}{ds^2} + \frac{\kappa^3}{2}\right) \quad (21)$$

and the Lagrange multiplier is identified with the regularized chemical potential (Eq.4). The critical shape, solution of the Euler-Lagrange, also satisfies

$$\kappa = 0 \quad as \quad x \to \pm\infty \quad (22)$$

$$\frac{\partial^2\kappa}{\partial s^2} = 0 \quad as \quad x \to \pm\infty \quad (23)$$

which implies $\lambda=0$. As a result, the saddle point associated with the nucleation barrier is characterized by the fact that the chemical potential is zero along the interface.

## 2. Critical shape and properties at the saddle-point

At the saddle point of the surface energy, the function $\theta(s)$ thus satisfies $\mu=0$ with the boundary conditions

$$\theta \to \theta_0 \quad as \quad s \to \pm\infty. \quad (24)$$

The non-dimensional problem can be written as

$$\Gamma^*(\theta)\frac{d\theta}{dS} - \left(\frac{d^3\theta}{dS^3} + \frac{1}{2}\left(\frac{d\theta}{dS}\right)^3\right) = 0 \quad (25)$$

$$\theta \to \theta_0 \quad as \quad S \to \pm\infty \quad (26)$$

where $S=s/L$ with a length scale defined as $L^2=\beta/\gamma_0$, $\gamma_0$ is the orientation-independent part of the interfacial energy and $\Gamma^*(\theta)$ the non-dimensional stiffness, $\Gamma(\theta)/\gamma_0$. As this is very similar to the semi-infinite wedge problem solved by Spencer [11] but with different boundary conditions, we follow the same procedure to determine the solution of Eq.25. It consists in treating $\theta$ as the independent variable and $K=d\theta/dS$ as the dependent one. This transforms the nonlinear third-order equation (Eq.25) into the linear problem,

$$\frac{d^2\varphi}{d\theta^2}+\varphi=\Gamma^*(\theta) \qquad (27)$$

with $\varphi=K^2/2$. The boundary conditions for $\varphi$ are trivial

$$\varphi=0 \quad at \quad \theta=\theta_0 \qquad (28)$$

$$\frac{d\varphi}{d\theta}=0 \quad at \quad \theta=\theta_0. \qquad (29)$$

The general solution of Eq.27 is given by

$$\varphi(\theta)=\gamma(\theta)+A\cos(\theta)+B\sin(\theta) \qquad (30)$$

where $A$ and $B$ are constants determined by the boundary conditions for the nucleation problem. From $\varphi$ the curvature is found

$$K=\pm\sqrt{2\varphi(\theta)}. \qquad (31)$$

This implies that $\varphi$ must be positive for $K$ to be real. Since the curvature, and thus φ, varies continuously along the critical shape, this determines entirely the solution. The function $\varphi$ is plotted in Fig.3 for two different values of the initial orientation $\theta_0= \theta^-$ and $\theta_0=$-0.675 rad, which respectively correspond to the stability limit and to a metastable orientation. As imposed by the boundary condition, $\varphi$ and $\varphi$' vanish when approaching $\theta_0$. For $\theta_0= \theta^-$, $\varphi$ increases with $\theta$, reaches a maximum in $\theta=0$ and decreases to $\varphi=0$ when $\theta$ approaches the upper stability limit $\theta=\theta^+$ (with $\varphi$' =0). This solution corresponds to the corner solution for the semi-infinite wedge derived by Spencer [11] and is plotted in Fig.4. For the metastable orientation, $\varphi$ first increases with $\theta$ when starting from $X<0$, this imposes $K>0$ and therefore leads to $K =+\sqrt{2\varphi(\theta)}$. For $X>0$, $\varphi$ increasing with $\theta$ means a negative curvature and thus $K = -\sqrt{2\varphi(\theta)}$. The two profiles can only connect where $\varphi=0$ as the curvature must evolve continuously. At this value of $\theta$ (~0.38 rad in this case), we arbitrary set $S=0$ (and $X=0$). Once $K(\theta)$ is known, the solution is obtained, integrating Eq.3,

$$S = \int_{\theta(S=0)}^{\theta_0} \frac{1}{K} d\theta. \quad (32)$$

Then $S(\theta)$ is inverted to find $\theta(S)$. As an illustration, the interfacial profiles for two different metastable orientations are given in Fig.5 for a given anisotropy function $\gamma = \gamma_0 (1+0.5\cos 4\theta)$. These are the critical shapes of the nucleation problem with corner energy regularization. Dimensionless lengths $X=x/L$ and $Y=y/L$ have been defined, where $L$ is the length scale introduced in Eq.25, $L = \sqrt{\beta/\gamma_0}$ with $\beta$ the regularization parameter and $\gamma_0$ the orientation-independent part of the interfacial energy. Therefore, the parameters $\beta$ and $\gamma_0$ control the size of the critical shape. From Eq.1, it immediately follows that the energy at the saddle point (i.e. the nucleation barrier) scales as $\sqrt{\beta\gamma_0}$; the factor of proportionality is computed numerically by inserting the critical shape into Eq.1. The nucleation barrier is shown in Fig.6. For a given $\gamma_0$ both the size of the critical shape and the energy of the saddle point increase with the regularization parameter $\beta$, both scaling as $\sqrt{\beta}$. However, the orientation at $S=0$, that we name the critical orientation by analogy with the classical nucleation theory, is independent of $\beta$ and $\gamma_0$ as those physical parameters do not appear in Eq.27, which controls the variation of the dimensionless curvature with orientation.

A close analogy is found between the problem of crystal surface formation and the Cahn Hilliard theory of nucleation. Indeed, the nucleation barrier is found to diverge when the initial orientation approaches the binodal limit and vanishes when reaching the spinodal regime, as in nucleation theory [25-29]. The size of the critical profile of the interface is also found to diverge near the binodal limit. When approaching the spinodal regime, the critical shape becomes a flat interface with an orientation that approaches the initial one. Both limits have their analogs in the Cahn and Hilliard theory [24].

The condition of zero chemical potential along the surface is found to require a motion of the initial surface for $X \to \pm\infty$ (Fig.5). This suggests that, in practice, nucleation of a stable orientation shall not operate through the saddle-point of the surface energy but potentially via ridge crossing through a more local change of the surface shape near the appearing stable orientation, even if this implies an increase of the nucleation barrier. This is further discussed in Sec. 5 below.

## 3. The Landau approach

Another regularization term has been proposed in previous works [22,23] and can be regarded as the simplest approximation, in the spirit of Landau, of the surface energy expansion in term of curvature. In this context, Eq.1 takes the following form

$$E = \int f(q) + \frac{\alpha}{2}\left(\frac{dq}{dx}\right)^2 dx \qquad (33)$$

with the projected energy on the *x*-axis *f(q)* and the slope *q* as defined in Eq.8 and Eq.9. In this expansion, only the first even term in the gradients has been retained for symmetry reasons and with a coefficient that is in principle a function of the slope but is regarded here as constant (as in [22,23]). This form of the energy is completely analogous to the one-dimensional version of the Cahn and Hilliard theory of diffuse interfaces where the slope plays the role of the order parameter and the function *f(q)* the thermodynamic potential. The regularization parameter $\alpha$ has the dimension of *β*. Minimization of the total energy is subject to the volume constraint given by Eq.17 and yields the following Euler-Lagrange equation

$$-\lambda - \frac{df_{y_x}}{dx} + \alpha y_{xxxx} = 0. \qquad (34)$$

This can be rewritten as

$$\lambda = \Gamma(\theta)\kappa + \alpha y_{xxxx} \qquad (35)$$

and as previously the Lagrange multiplier is identified with an alternative formulation of the regularized chemical potential and thus $\lambda=0$. Rewriting Eq.35 in terms of the slope and integrating with respect to $x$ gives

$$\alpha q_{xx} = \Delta f_q \qquad (36)$$

where $\Delta f_q$ is the driving force for nucleation

$$\Delta f_q = f_q - f_q\big|_{q=q_0} \qquad (37)$$

with $q_0$ the initial slope. Defining $X=x/L$ with a length scale $L^2=\alpha/\gamma_0$ and the non-dimensional driving force $\Delta f_q^* = \Delta f_q / \gamma_0$, we obtain

$$q_{XX} = \Delta f_q^*. \qquad (38)$$

For comparison with the previous regularization, an interfacial profile satisfying Eq.38 for a given initial slope is shown in Fig.5. The shape is very similar to the one with the curvature-squared regularization, and the properties at the saddle point of the surface energy such as the size of the newly formed crystal surface, its orientation, and the corresponding nucleation barrier show the same trends, as expected from the strict analogy revealed by Eq.38 between the nucleation problem of crystal surfaces and the non-classical theory.

## IV. EQUATION OF MOTION

Once the critical interfacial shape is formed, the system enters the growth regime. Growth can be driven by various mechanisms, such as condensation-evaporation (or motion by curvature), deposition or surface diffusion. We only present the condensation-evaporation mechanism for which the rate of evaporation or condensation is proportional to the difference in chemical potential from equilibrium (zero for an infinite plane) and the surface area [22,30,31]:

$$\frac{\partial y}{\partial t} = -A\sqrt{1+q^2}\,\mu \quad (39)$$

with *A* a positive kinetic constant related to the frequency with which atoms attach to and detach from the surface. Eq.39 leads to a deterministic equation for the slope,

$$\frac{\partial q}{\partial t} = A\frac{\partial}{\partial x}\left(\sqrt{1+q^2}\,\mu\right). \quad (40)$$

We linearize the equation of motion (Eq.39) about an initial and uniform state for the curvature-squared regularization and the corresponding form of the chemical potential (Eq.4). Let,

$$y = -q_0 x + \delta y \exp(i\mathbf{k}\cdot\mathbf{x} + \omega t) \quad (40)$$

where $q_0$ specifies the initial orientation, $\delta y$ is a small constant and $\omega(\mathbf{k})$ is the growth rate. Substituting Eq.40 into the equation of motion and keeping only the terms linear in $\delta y$, we find from the dispersion relation that the equation of motion exhibits a marginally critical stable wave vector $k_c$ where $\omega(k_c)=0$,

$$k_c^2(\theta) = -\frac{\Gamma(\theta_0)}{\beta\cos^4\theta} = -\frac{1}{L^2}\frac{\Gamma^*(\theta_0)}{\cos^4\theta} \quad (41)$$

and, as in the Cahn-Hilliard theory, the initially flat interface with $\Gamma(\theta_0)<0$ is unstable and will separate into domains of characteristic size given by ~ $1/k_c$. Such an analogy with the Cahn-Hilliard analysis is not surprising since it has been formally shown that the long-wave approximation for the anisotropic motion by curvature gives a convective Cahn-Hilliard equation for the slope [13,14], and a higher-order convective Cahn-Hilliard equation for surface diffusion [15]. A detailed review of the different coarsening regimes with deposition or constant driving force and kinetic effects is given in ref.20, where the combined effects of all the processes have been investigated and the long-wave approximation performed. The analogy between "faceting" [32,33] and the theory of spinodal decomposition is straightforward when considering the regularization proposed in Eq.33. Indeed, assuming that $|q|<<1$ the equation for the slope is

$$\frac{\partial q}{\partial t} = A \frac{\partial^2}{\partial x^2}\left(-\alpha q_{xx} + f_q\right). \quad (41)$$

This is identically the Cahn-Hilliard equation describing spinodal decomposition of a binary system, where the slope plays the role of the composition. Consequently, the dynamics of crystal surfaces, in terms of slope, strictly follows that of the composition in the Cahn-Hilliard equation.

## V. SADDLE POINT NUCLEATION AND RIDGE CROSSING

To validate the theoretical predictions developed in the preceding section, we begin with a numerical test of the saddle point solution. To this end, we use a dimensionless form of Eq.39 where lengths are given in units of $L$ and time $t^*$ in units of $L^2/A\gamma_0$. This equation is discretized on a regular grid with constant spacing $\Delta X$ by standard finite-difference formulas, and integrated in time with an explicit Euler scheme of timestep $\Delta t^*$. Since the critical shape satisfies $\mu = 0$ everywhere, it is a stationary solution of Eq.39, as expected. We initialize the simulations with shapes that are slightly post-critical or pre-critical which are constructed from the saddle-point solution: a patch of the new stable orientation is inserted, with the orientation given by the analytical solution but a length which is slightly above or below the critical length (defined for example by the zeros of $q_{ss}$). For convenience, we rotate the Wulff plot such as to align the initial orientations $\theta_0$ with the $x$-axis. This modifies the potential of interest $f(q)$ and shifts the angles $\theta^-$ and $\theta^+$, as well as the location of the spinodal, by $-\theta_0$. Results are shown Fig.7 for $q_0=-0.8$ (which is the same value as for the example shown in Fig.5). As expected, for the pre-critical shape, the pseudo-facet shrinks and ultimately disappears, and the surface relaxes back to the metastable planar interface. For the post-critical shape, the patch of the stable orientation grows in size and is surrounded by regions of the other stable orientation, which, in this case, is very close to the orientation of the initial metastable surface. These results demonstrate that the critical shape is at the border of two different regimes (growth and disappearance of the new orientations), as expected from a saddle-point solution.

The saddle-point solution exhibits finite shifts of the surface height for $X \to \pm\infty$ (Fig.5), that is, non-local changes are necessary to reach the saddle point from the initial metastable interface. For a random evolution driven by uncorrelated local fluctuations, such an event is highly unlikely. Nucleation should rather proceed by a more local deformation of the surface. To test this scenario, we simulate the dynamic evolution of a metastable initial flat surface using Eq.39 to which a small noise is added. This is done by adding a noise term of the form $a\chi\sqrt{1+q^2}$ to the dimensionless evolution equation for the surface height, with $\chi$ uniformly distributed on the interval [-1,1], $\sqrt{1+q^2}$ accounts for the surface geometry, and $a$ a dimensionless amplitude. As is well known from statistical field theory [34], such a Langevin term corresponds to thermal fluctuations with a temperature that is proportional to $a^2$. The precise relationship between the temperature and the numerical parameters also depends on the spatial and temporal discretization [34,35]. Moreover, it is known that non-linear equations such as Eq.39 are renormalized by noise, and only in the limit of small noise the calculation of the critical shape performed without fluctuations remains a good approximation [35]. Since we are not aiming for a quantitative comparison of theory and simulations, but rather want to investigate qualitatively the nucleation pathway, we do not need to establish a precise expression for the temperature; however, we need to remain in the small-noise regime. The maximum value of the noise that we have used in the dynamical simulations is $a\Delta t^* = 2,5\times 10^{-3}$. It will be shown below that this is indeed small enough.

The time evolution of a metastable surface is shown Fig.8 for $q_0=-0.42$. In total, five patches of the stable orientation nucleate spontaneously over the duration of the simulation. Nucleation proceeds by local deformations of the initial surface due to the fluctuations. This is clearly visible at $t^*=10$ for three nucleation events. The two next nucleation events occur between $t^*=10$ and $t^*=100$. Then, the growth and coarsening regime sets in; one pseudo-facet disappears around $X=15$ at longer times (see Fig.9). Coarsening is very slow, as also observed in previous works, such as in ref.12 where the morphological evolution is driven by surface diffusion. The surface profiles are very similar to the ones obtained by phase-field simulations of spinodal faceting with surface diffusion [36,37]. This is quite natural since the methods used to regularize the problem are formally different, but also give an energy and a length scale to corners. Moreover, it is anticipated that if the evaporation-condensation mechanism is replaced by surface diffusion, the emerging morphologies will remain approximately the same in the nucleation regime, since the critical shape is identical for the two dynamics.

It can be checked that the nucleation rate is compatible with classical nucleation theory. To this end, we perform simulations with different noise strength and count the number of nucleation events after a fixed time. The resulting data are plotted in Fig. 9 versus $a^2$ and compared to the simplest expression for the nucleation rate, $N = k_1 e^{-k_2/a^2 \Delta t^*}$ where $k_1$ and $k_2$ are constants. The fit is reasonable and yields a dimensionless nucleation barrier that is compatible with the one calculated from the saddle-point solution. Moreover, the pre-factor of the nucleation rate is also compatible with classical nucleation theory, as the fit gives a value of 0.07 $L^{-1}$, which is close the possible nucleation sites density ~0.1 $L^{-1}$, evaluated as the typical size of the observed nucleus divided by the system size (Fig.8). This confirms that the noise is small enough to neglect the renormalization of the evolution equation.

This suggests that the most likely pathway for nucleation does not pass through the saddle point of the surface energy. Instead, the system crosses the "ridge" that separates the basins of attraction of the metastable and stable states at a different point which corresponds to a more local deformation of the crystalline surface. The central region of the nucleated patches has a geometry that is very close to the saddle-point solution; therefore, the chemical potential is different from zero only in the smooth "junctions" between the newly formed patch and the original surface. Thus, while the nucleation barrier is necessarily higher than for saddle point nucleation, the difference between the two values is likely to be small.

## VI. CONCLUSION

By deriving the Euler-Lagrange equation for crystal surface nucleation with corner energy regularization in two dimensions, we have shown that the critical surface corresponds to the one for which the chemical potential vanishes. We have then examined the properties of the critical shape, and we have shown that the problem of formation of stable orientations is analogous to the diffuse-interface theory of nucleation, with a vanishing nucleation barrier near the spinodal limit. We have demonstrated that the problem of crystal surface formation is strictly identical to the Cahn-Hilliard theory of nucleation in a binary system when regularization is made in the spirit of Landau. In this case, the evolution equation for motion by curvature is a Cahn-Hilliard equation for the slope. The linear stability analysis of the equation of motion for the curvature-squared regularization also strengthens this analogy. Finally, it was shown by direct simulations of the equation of motion that ridge-crossing is preferred over saddle-point nucleation since the latter would require a non-local deformation of the initial surface.

**Acknowledgments.** We acknowledge T. Pusztai (Wigner Research Centre for Physics) for helpful discussions.

**Funding.** This work was supported by the ANR ANPHASES project (M-ERA.NET).

**Figure captions**

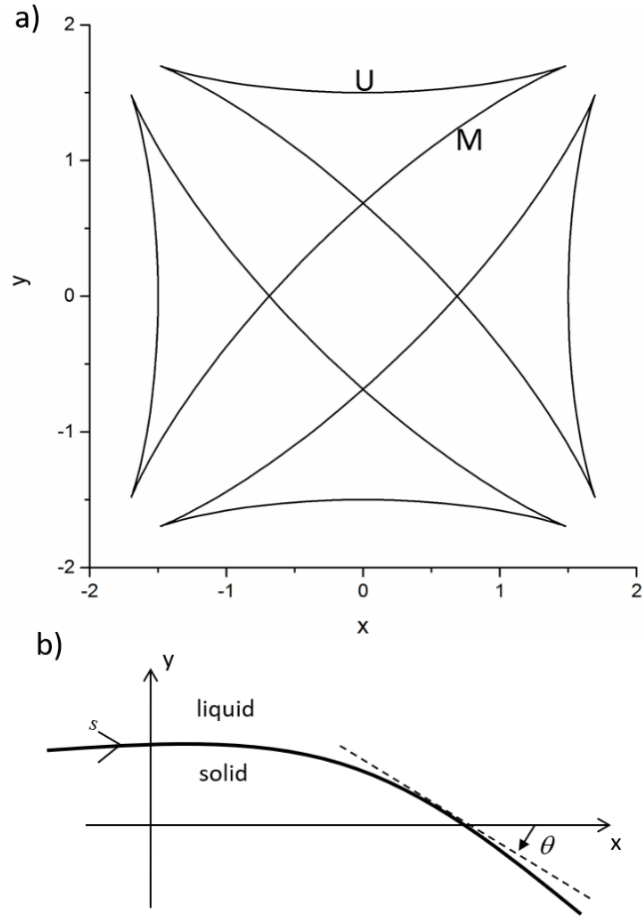

Fig.1 a) The ξ-plot for $\gamma = 1 + 0.5\cos 4\theta$. The interior part of the ξ-plot gives the Wulff shape. In region U orientations are unstable and have a negative stiffness ($\gamma + \gamma_{\theta\theta} < 0$) and orientations in region M are metastable. b) Schematic representation of the solid surface, which is described by a curve in the (*x*,*y*) plane and parametrized by an arclength *s* chosen so that when travelling in the increasing s direction along the interface, the solid is on the right. The orientation of the interface is given by the tangent angle to the interface $\theta$ measured clockwise from the *x*-axis.

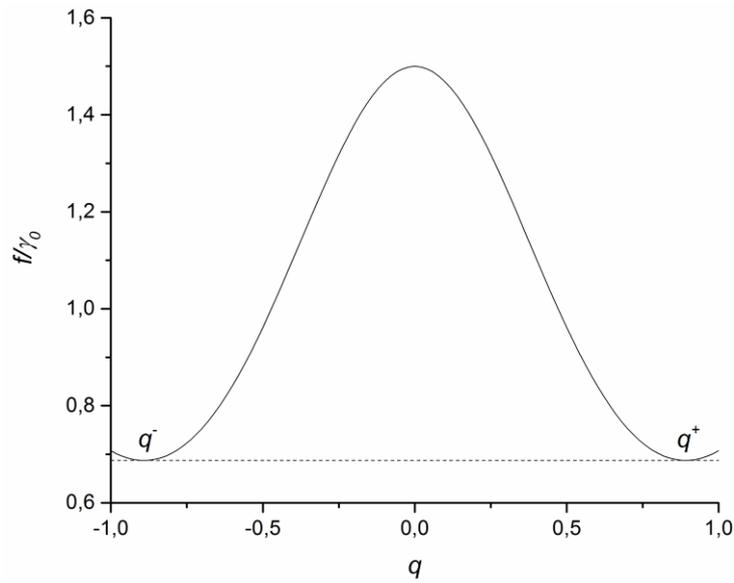

Fig.2 The double well function $f$, defined as $\gamma/\cos\theta$, for the following anisotropy function $\gamma = \gamma_0(1+0.5\cos 4\theta)$.

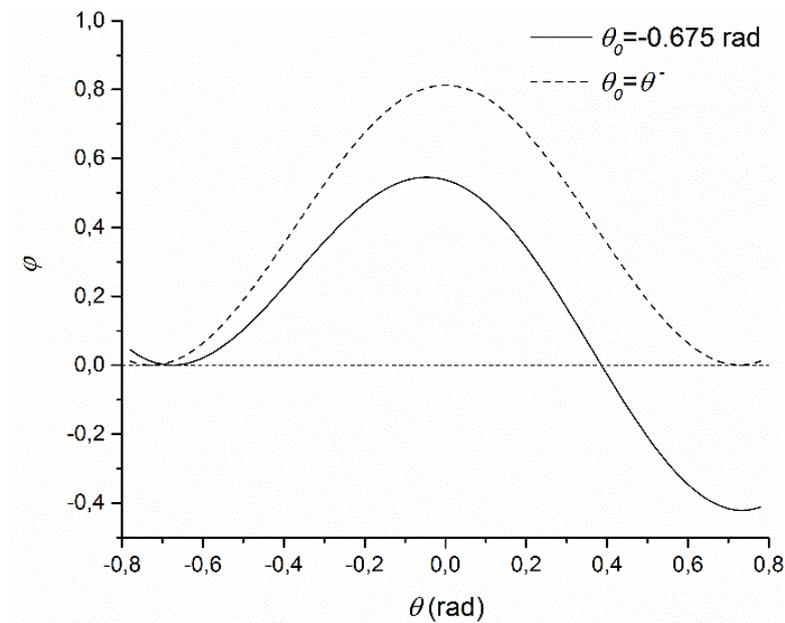

Fig.3 The solution of Eq.27 for two different initial orientations, which obeys the boundary conditions given by Eq.28 and 29. $\varphi$ is defined as $\varphi = K^2/2$ with $K$ the curvature.

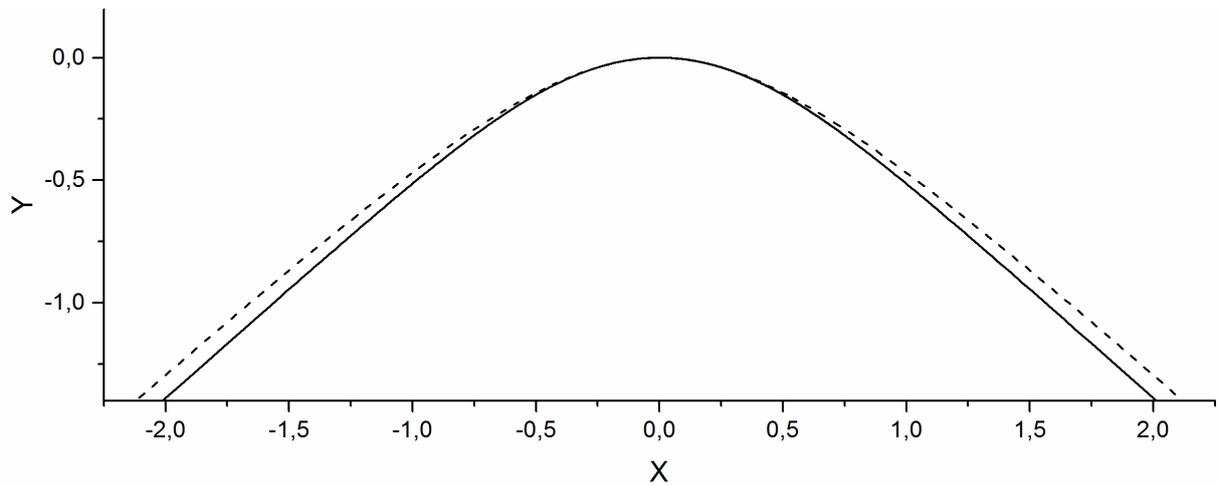

Fig.4 The semi-infinite wedge solution for the curvature-squared regularization (full line) and the Landau approximation (dashed line). $X=x/L$ and $Y=y/L$.

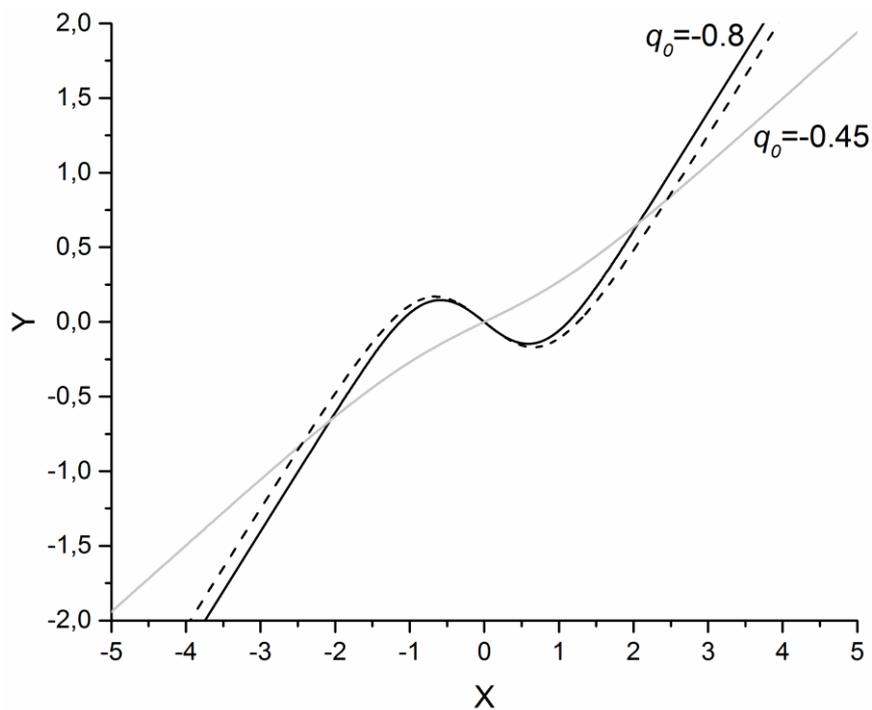

Fig.5 Crystal shapes at the saddle point of the energy for different initial slopes. The dashed-line is used for the Landau regularization. $X=x/L$ and $Y=y/L$.

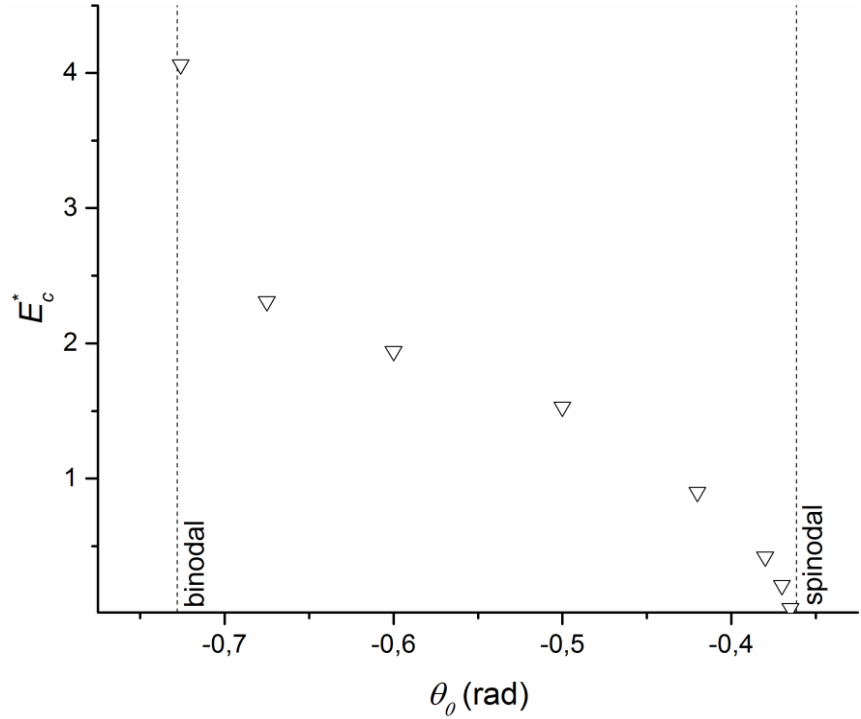

Fig.6 Energy at the saddle point (i.e. the nucleation barrier), divided by $\sqrt{\beta\gamma_0}$, as a function of the initial orientation.

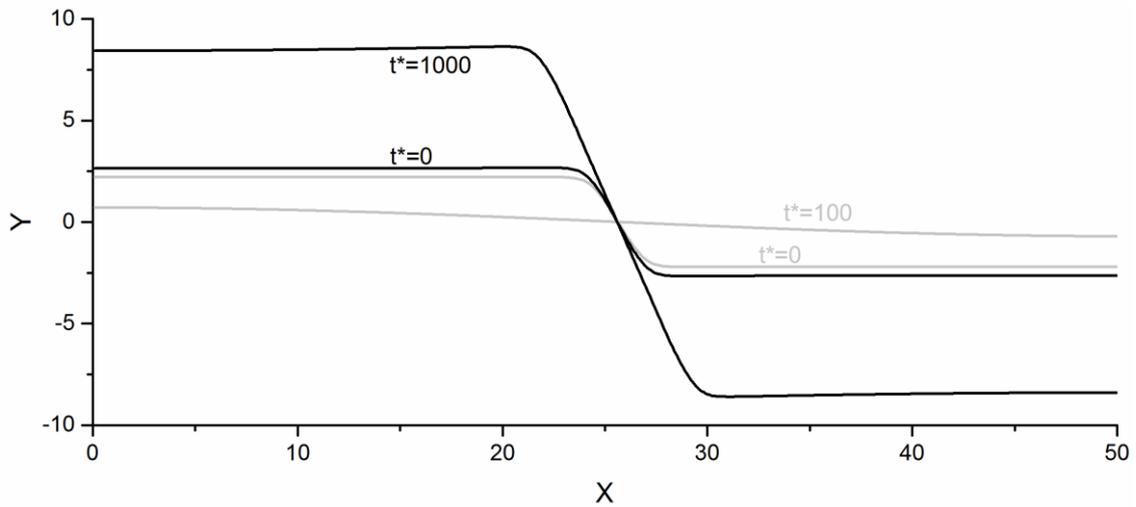

Fig.7 Dynamic evolution of crystalline surfaces for two initial shapes, in grey for the pre-critical shape and in black for the post-critical one. The two initial surfaces are very close, the length of the appearing facet is slightly below its critical value for the pre-critical shape and slightly above for the post-critical one. $q_0=-0.8$, $X=x/L$ and $Y=y/L$. Time is given in units of $L^2/A\gamma_0$. Landau regularization and no-flux boundary conditions are used.

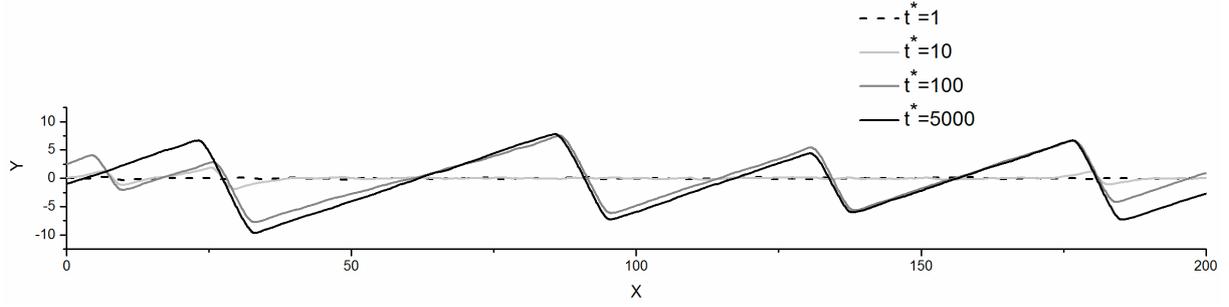

Fig.8 Dynamic evolution of a metastable initial surface. A small noise is added on the interface so that to promote phase separation by adding to Eq.39 a term in the form $a\chi\sqrt{1+q^2}$, with $\chi$ uniformly distributed on the interval [-1,1]. $q_0$=-0.42, X=x/L, Y=y/L and $a\Delta t^* = 2,5\times 10^{-3}$. Time is given in units of $L^2/A\gamma_0$. Landau regularization and periodic boundary conditions are used.

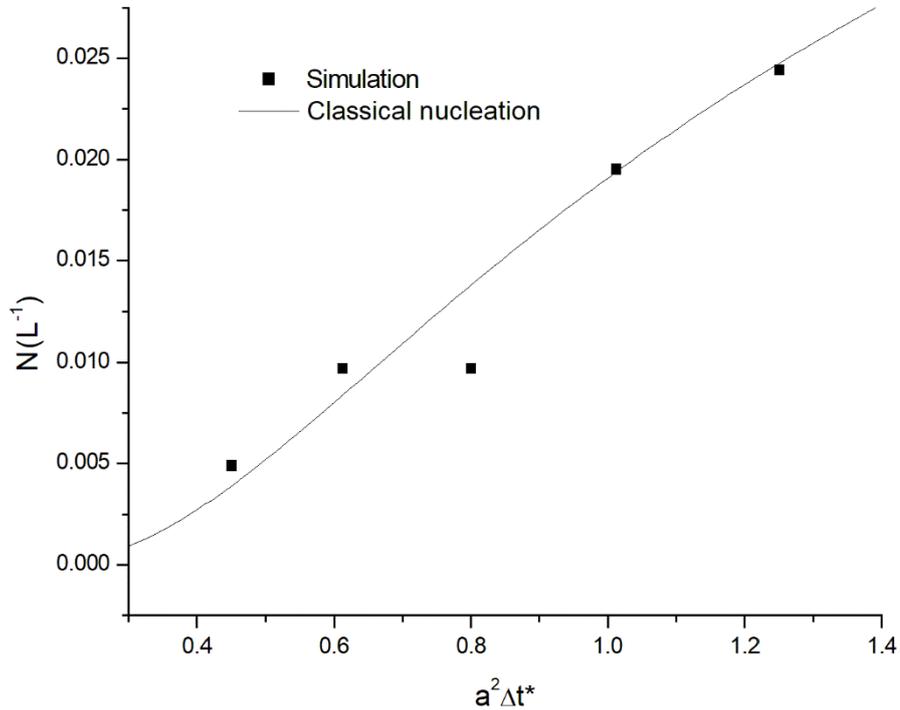

Fig.9 Number density (N) of nucleated surfaces as a function of $a^2\Delta t^*$; which is proportional to the temperature. Simulations are compared to the simplest classical nucleation model, $N = k_1 e^{-k_2/a^2\Delta t^*}$ where $k_1$ and $k_2$ are constants, respectively equal to 0.07 $L^{-1}$ and 1.3.